\documentclass{mem}
\usepackage{natbib}\usepackage{txfonts}\usepackage{balance}
\usepackage{graphicx}
\usepackage[a4paper,breaklinks,dvipdfm]{hyperref}
\idline{75}{282}
\begin{document}
\def\teff{$T\rm_{eff }$}
\def\kms{$\mathrm {km s}^{-1}$}

\title{
Accretion and outflow in the gamma ray bursts from black hole binary systems
}

   \subtitle{}

\author{
A. \,Janiuk\inst{1}, S. Charzy\'nski\inst{2}
\and P. \,Mioduszewski\inst{1}
          }

  \offprints{A. Janiuk}

\institute{Center for Theoretical Physics, Polish Academy of Sciences, 
Al. Lotnikow 32/46, 02-668 Warsaw, Poland\email{agnes@cft.edu.pl}
\and
Faculty of Mathematics and Natural Sciences, Card. Stefan Wyszynski University, ul Dewajtis 5, 01-815 Warsaw, Poland
}

\authorrunning{Janiuk}

\titlerunning{GRBs central engine}

\abstract{

We consider a scenario for the longest duration gamma ray bursts, resulting from the collapse of a massive star in a close binary system with a companion 
black hole.
The primary black hole born during the core collapse 
is spun up and increases its mass during the fallback of the stellar envelope.
 The companion black hole provides an additional angular momentum to the 
envelope, which ultimately makes the core BH spinning with a high Kerr parameter. 
After the infall and spiral-in, the two black holes merge inside the 
circumbinary disk. The second episode of mass accretion and final, even larger
 spin of the 
post-merger black hole prolongs the gamma ray burst central engine activity. 
The observed events should have two distinct peaks in the electromagnetic 
signal, separated by the gravitational wave emission. The gravitational recoil
of the burst engine is also possible.

\keywords{Accretion, accretion disks -- Black hole physics -- 
Gamma ray bursts: general -- Stars: massive }
}
\maketitle{}



The first stage of our simulation is the collapse of a 
massive star envelope onto the core black hole, taken as in \citet{janiuk08}. 
The rotating inner shells accrete prior to the companion infall, changing 
the mass and spin
of the primary black hole, and particular results depend on the magnitude of 
specific angular momentum in the envelope.
The additional spin-up is due to the companion black hole 
orbiting in the binary system and entering the outer envelope \citep{bk10}. 
We allow also that a 
fraction (up to $\sim 30\%$) of the envelope's mass is lost by the wind. 
This stage lasts about 500-700 seconds, and 
as the inner disk has been accreted, 
the binary black hole merger
occurs in the vacuum gap, before the outer circumbinary disk falls 
onto the merger product \citep{farris11}. 
We consider non-equal masses, large spin of the primary and zero spin of the 
secondary black holes. The spin vector is perpendicular to the orbital plane.
The merger timescale is only about $\Delta t_{\rm m}\sim 2.5\times 10^{-3}$ 
seconds. The gravitational recoil occurs with a large velocity of
$v_{\rm rec}\approx 10^{5}$ km s$^{-1}$. The simulation is ended after about 
0.01 s, when the new apparent horizon of the product black hole is found.

The third, final stage of the GRB engine activity is the accretion of outer
disk onto the product: spinning, massive black hole.
We consider the material gravitationally bound to the recoiled black hole and 
we compute the relativistic MHD model \citep{gammie03}, parameterized with
the black hole mass, spin and mass of the bound torus, with values
 estimated from the results of the former modelling. 
 The magnetized plasma in the torus is cooled by neutrinos which are formed in
the nuclear reactions at high temperatures and densities \citep{janiuk07}.
 During the time evolution, the total energy is reduced after each time-step to account for neutrino cooling.
 Neutrino luminosity of the engine is dominated by the hot winds launched at medium latitudes from the torus surface.


\begin{figure}[]
\resizebox{.8\hsize}{!}{\includegraphics[clip=true]{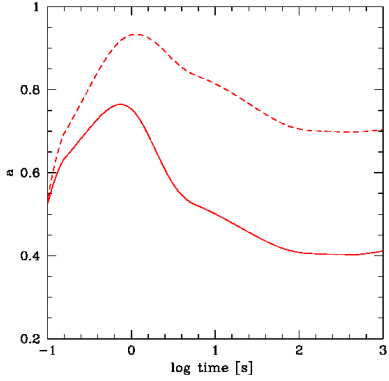}}
\caption{
\footnotesize
Spin of the primary core black hole as a function of time during the
massive star collapse. 
The two lines show the models with different magnitude 
of the specific angular momentum in the envelope: $l_{\rm spec}/l_{\rm crit}=1.5$ 
(solid), or
$l_{\rm spec}/l_{\rm crit}=3.0$ (dashed). The value of $l_{\rm crit}$ gives 
the condition for the formation of accretion disk. Initial spin of $a=0.5$ is assumed. Black hole mass changes from 1.7 to 9.2 $M_{\odot}$ (no wind loss in this model). 
}
\label{fig:primary_spin}
\end{figure}


\begin{figure}[]
\resizebox{\hsize}{!}{\includegraphics[clip=true]{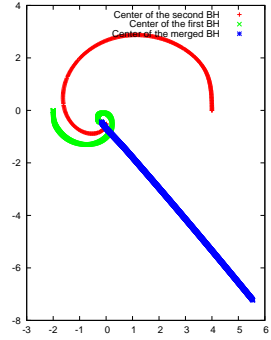}}
\caption{
\footnotesize
The final stage of binary black hole merger. Lines show the trajectories of the components and product black holes in x-y plane (dimensionless units).
The mass ratio of the black holes is 2.5. 
The spin of the heavier black hole is $a_{1}=0.68$
 and spin of the secondary black hole is $a_{2}=0$. The product black hole 
obtained a spin of 0.76 and gravitational recoil kick.
}
\label{fig:bbh_merger}
\end{figure}






The spinning black hole and strong magnetization of the accreting plasma 
allows to launch the bipolar jets.
Neutrinos emitted in the torus winds may annihilate and provide additional source of power.
The Blandford-Znajek luminosity of the jets and estimated 
annihillation luminosity are equal to $L_{\rm BZ}= 2.96 \times 10^{52}$ 
and $L_{\rm \nu}= 2.34 \times 10^{54}$ erg s$^{-1}$ at the end of 
our sample simulation. The parameters used here: $a=0.7$, $M_{\rm BH}=11.5 M_{\odot}$
 and initial $M_{\rm torus} = 7 M_{\odot}$. The latter is about 45\% of the remaining envelope mass. Further 3.5 $M_{\odot}$ was either accreted or 
lost through the wind during 0.1 s of this MHD simulation.

To sum up, we considered a long GRB scenario, that consists of 3 steps: (i) long, 
moderately energetic GRB powered by the inner torus accretion onto the core black hole within the massive star's envelope, spun up by the infalling companion
 (ii) binary black hole merger with product recoil and (iii) fast, energetic jet
launched by MHD mechanism via accretion onto the 
spinning black hole, and powered additionally by neutrino annihillation.

\begin{acknowledgements}
This research was supported in part by grant NN 203 512638 
from the Polish Ministry of Science and Higher Education.
\end{acknowledgements}

\bibliographystyle{aa}

\end{document}